\documentclass[aoas,preprint]{imsart}

\RequirePackage[OT1]{fontenc}
\RequirePackage[numbers]{natbib}

\usepackage{amsthm,amsmath}
\usepackage{amsbsy}

\input xy
\xyoption{all}

\usepackage{algorithm}
\usepackage[noend]{algpseudocode}

\usepackage{epsf}
\usepackage{epsfig}

\startlocaldefs
\newcommand{\cov}{\mbox{Cov}}
\newcommand{\bfU}{\boldsymbol{U}}

\newcommand{\bfL}{\boldsymbol{L}}
\newcommand{\bfH}{\boldsymbol{H}}

\newcommand{\ind}{1\!\!1}
\newcommand{\ci}{\perp\!\!\!\perp}

\newcommand{\bfC}{\boldsymbol{C}}

\newcommand{\bfV}{\boldsymbol{V}}

\newcommand{\bfmH}{\boldsymbol{\mathcal{H}}}
\newcommand{\bfmV}{\boldsymbol{\mathcal{V}}}

\floatstyle{plain}
\newfloat{supplefig}{htbp}{lgr}
\floatname{supplefig}{Figure S}

\newfloat{suppletextfig}{htbp}{lgr}
\floatname{suppletextfig}{Figure ST}

\endlocaldefs

\begin{document}

\begin{frontmatter}

\title{Using instrumental variables to disentangle treatment and placebo effects in blinded and unblinded randomized clinical trials influenced by unmeasured confounders}
\runtitle{IVs for disentangling medication and placebo effects}


\author{\fnms{Elias} \snm{Chaibub Neto \\ \footnotesize{Sage Bionetworks}}\ead[label=e1]{elias.chaibub.neto@sagebase.org, Sage Bionetworks}}
\address{\printead{e1}}


\runauthor{Chaibub Neto E.}

\begin{abstract}
Clinical trials traditionally employ blinding as a design mechanism to reduce the influence of placebo effects. In practice, however, it can be difficult or impossible to blind study participants and unblinded trials are common in medical research. Here we show how instrumental variables can be used to quantify and disentangle treatment and placebo effects in randomized clinical trials comparing control and active treatments in the presence of confounders. The key idea is to use randomization to separately manipulate treatment assignment and psychological encouragement messages that increase the participants' desire for improved symptoms. The proposed approach is able to improve the estimation of treatment effects in blinded studies and, most importantly, opens the doors to account for placebo effects in unblinded trials.
\end{abstract}





\end{frontmatter}

\section{Introduction}

Placebo effects have draw a lot of interest and debate in medicine\cite{price2008}. They can be viewed as a simulation of an active therapy within a psychosocial context\cite{price2008}. Research in neurobiology has shown that placebo responses are accompanied by actual alterations in neural activity within brain regions involved in emotional regulation\cite{price2008,fields2004,petrovic2005,zubieta2001}. Hence, rather than inducing a simple bias in response, placebos can induce actual biological effects and improve clinical outcomes. Among the cognitive and emotional factors that have been proposed to contribute to placebo effects, the interaction between the desire for symptom change and the expected symptom intensity has been proposed as a key component giving raise to placebo effects\cite{price2008}. In the psychology literature, this interaction is known as the desire-expectation model of emotions\cite{price2008,price1984,price1985,price2001}, which postulates that ratings of positive and negative emotional feelings are predicted by multiplicative interactions between ratings of desire and expectation. A number of experimental studies of placebo analgesia\cite{price2008,vase2003,verne2003} have corroborated the role of the desire-expectation model as a trigger of placebo effects. These findings have important implications for both clinical practice and clinical trials. On one hand, clinicians should harness the placebo effect to improve the clinical outcome of their patients (by managing expectations and desires through ethical use of suggestions and optimum caregiver-patient interactions)\cite{price2008}. On the other hand, assessment of expectation and desire levels is also important in clinical trials since placebo effects might strongly influence the results of a study. In unblinded trials, it is widely recognized that the overall effect attributed to a treatment might actually correspond to a combination of treatment and placebo effects. However, placebo effects might still play a role in blinded trials as well\cite{price2008}. For instance, blinded studies evaluating the effectiveness of acupuncture\cite{bausell2005} and of implantation of human embryonic dopamine neurons into the brains of persons with severe Parkinson disease\cite{mcrae2004} have shown that perceived treatment (or the treatment the participants thought they had received) can have stronger effects than the treatment actually received by the participants. These findings illustrate the relevance of measuring expectation, desire, and emotional levels in order to assess the contribution of placebo effects, and suggest that it is important to adjust for these variables when estimating treatment effects and interpreting the results of clinical trials\cite{price2008}. However, because it is generally impossible to rule out the presence of unmeasured confounders, simply measuring and adjusting for variables associated with placebo effects might not be enough to ensure a reliable estimation of the treatment effect. For instance, estimation based on regression models adjusting for the placebo related measurements still leads to biased estimates of the treatment effect, unless all confounders influencing the outcome variable enter the regression model.

\section{The statistical method}

Here we present a statistical approach to disentangle treatment and placebo effects using instrumental variables\cite{angrist2001,bowden1990,didelez2010} in randomized experiments. An instrumental variable (IV) is statistically independent from any unmeasured confounders, but is associated with the treatment variable and with the outcome variable (via its influence on the treatment variable alone). Use of IVs in randomized experiments allows the consistent estimation of treatment effects without the need to explicitly model the confounders (the technique even accounts for confounders the researcher is unaware about).

Our proposed method requires the ability to assess variables associated with placebo effects (e.g., levels of expectancy, desire, and emotion), and uses randomization to separately manipulate a pair of variables. The first, corresponds to a psychological encouragement variable aiming to increase the desire for improved symptoms. The study participants are randomized according to whether they receive the psychological encouragement or not. This ``psychological treatment" IV allows the consistent estimation of the placebo effect on the outcome in the presence of confounders. The second, corresponds to a treatment assignment variable representing the random assignment of participants to active treatment or control therapy groups. It allows the estimation of the treatment effect on the outcome, after adjustment for the placebo effect. Mechanistically, the approach corresponds to a two-step procedure, which first estimates the contribution of the placebo effect on the outcome, and then the effect of the treatment on the residuals of the outcome variable after the contribution of the placebo effect has been removed.

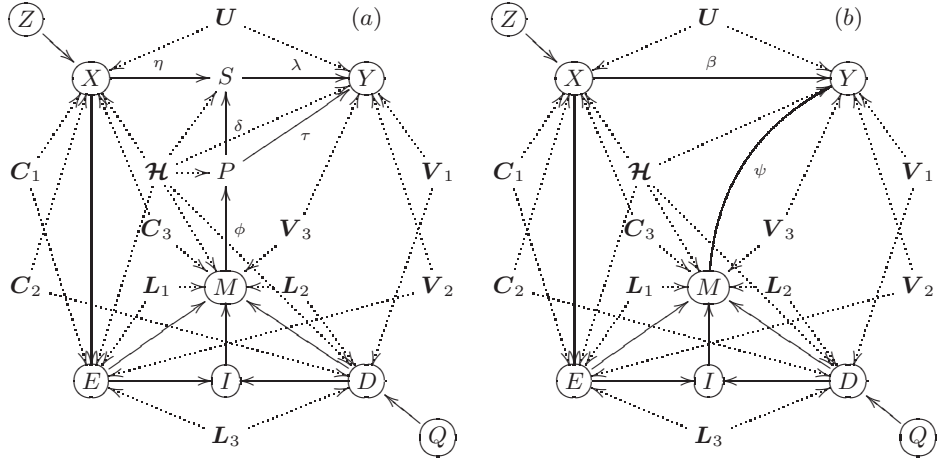
\begin{figure}[!h]
$$
\xymatrix@-1.3pc{
*+[F-:<10pt>]{Z} \ar[dr] &&& \bfU \ar@{.>}[dll] \ar@{.>}[drr] && (a) & &  *+[F-:<10pt>]{Z} \ar[dr] &&& \bfU \ar@{.>}[dll] \ar@{.>}[drr] && (b) \\
 & *+[F-:<10pt>]{X} \ar[rr]^{\eta} \ar[dddddd] && S \ar[rr]^{\lambda} && *+[F-:<10pt>]{Y} & &  & *+[F-:<10pt>]{X} \ar[rrrr]^{\beta} \ar[dddddd] && && *+[F-:<10pt>]{Y} & \\
 &&&&&&  &  &&&&&& \\
\bfC_1 \ar@{.>}[uur] \ar@{.>}[ddddr] && \bfmH \ar@{.>}[r] \ar@{.>}[ddr] \ar@{.>}[uur] \ar@{.>}[uul] \ar@{.>}[uurrr] \ar@{.>}[ddddl] \ar@{.>}[ddddrrr] & P \ar[rruu]_{\tau} \ar[uu]_{\delta} &&& \bfV_1 \ar@{.>}[uul] \ar@{.>}[ddddl] & \bfC_1 \ar@{.>}[uur] \ar@{.>}[ddddr] && \bfmH \ar@{.>}[ddr] \ar@{.>}[uul] \ar@{.>}[uurrr] \ar@{.>}[ddddl] \ar@{.>}[ddddrrr] &  &&& \bfV_1 \ar@{.>}[uul] \ar@{.>}[ddddl] \\
 && \bfC_3 \ar@{.>}[dr] \ar@{.>}[uuul] && \bfV_3 \ar@{.>}[dl] \ar@{.>}[uuur] &&  &  && \bfC_3 \ar@{.>}[dr] \ar@{.>}[uuul] && \bfV_3 \ar@{.>}[dl] \ar@{.>}[uuur]  &&   \\
\bfC_2 \ar@{.>}[ddrrrrr] \ar@{.>}[uuuur] && \bfL_1 \ar@{.>}[ddl] \ar@{.>}[r] & *+[F-:<10pt>]{M} \ar[uu]_{\phi} & \bfL_2 \ar@{.>}[ddr] \ar@{.>}[l] && \bfV_2 \ar@{.>}[uuuul] \ar@{.>}[ddlllll] &  \bfC_2 \ar@{.>}[ddrrrrr] \ar@{.>}[uuuur] && \bfL_1 \ar@{.>}[ddl] \ar@{.>}[r] & *+[F-:<10pt>]{M} \ar@/^1.25pc/[uuuurr]_{\psi} & \bfL_2 \ar@{.>}[ddr] \ar@{.>}[l] &&  \bfV_2 \ar@{.>}[uuuul] \ar@{.>}[ddlllll] & \\
 &&&&&& &  &&&&&& \\
 & *+[F-:<10pt>]{E} \ar[rruu] \ar[rr] && *+[F-:<10pt>]{I} \ar[uu] && *+[F-:<10pt>]{D} \ar[lluu] \ar[ll] & &   & *+[F-:<10pt>]{E} \ar[rruu] \ar[rr] && *+[F-:<10pt>]{I} \ar[uu] && *+[F-:<10pt>]{D} \ar[lluu] \ar[ll] & &   \\
&&& \bfL_3 \ar@{.>}[ull] \ar@{.>}[urr] &&& *+[F-:<10pt>]{Q} \ar[lu] & &&& \bfL_3 \ar@{.>}[ull] \ar@{.>}[urr] &&& *+[F-:<10pt>]{Q} \ar[lu]  \\
}
$$
\caption{Direct acyclic graph representation of the causal model underlying the proposed IV approach for disentangling treatment and placebo effects in unblinded clinical trials. Circled and un-circled nodes represent observed and unobserved variables, respectively. Arrows represent the causal influence of a variable on another, with the influence of confounders on variables shown as dotted arrows. The $Z$ and $X$ nodes represent, respectively, the participant's assigned and received treatment, whereas $Q$ stands for the psychological encouragement treatment. The $S$ and $P$ variables represent the (unobserved) somatic and psychosomatic states of the participant, respectively. The $E$, $D$, $I$, and $M$ nodes stand for the participant's expectation of symptom intensity, desire for improved symptoms, desire-expectation interaction, and emotional level, respectively. The sets of variables $\bfU$, $\bfC_1$, $\bfC_2$, $\bfC_3$, $\bfL_1$, $\bfL_2$, $\bfL_3$, $\bfV_1$, $\bfV_2$, $\bfV_3$, and $\bfmH$ stand for unmeasured confounder variables. The $Y$ node represents the outcome variable. Panel a shows the full model. Panel b shows the reduced model where the unobserved somatic and psychosomatic states of a participant are not directly represented in the causal model.}
\label{fig:dags}
\end{figure}

A graphical representation of the causal model underlying our approach is given in Figure \ref{fig:dags}a. Circled and un-circled nodes represent observed and unobserved variables, respectively. Arrows represent the causal influence of a variable on another, with the influence of unmeasured confounders shown as dotted arrows. The binary variable $Z$ represents the randomized treatment assigned to the participant (1 if participant is assigned to the active treatment group, and 0 if assigned to the control group), while $X$ represents the treatment actually received by the study participant (1 if the participant receives the active treatment, and 0 otherwise). It is important to model both assigned and received treatment variables since participants won't necessarily subscribe to their assigned treatment, and the experiment might suffer from imperfect compliance.

The variable $S$ represents the unmeasured biochemical/physiological (somatic) state of a participant and mediates the effect of the treatment on the outcome variable, $Y$. For instance, if $X$ represents a drug treatment, then $S$ could represent the physiological state induced by the biochemical pathways targeted by the drug. The causal effects of $X$ on $S$ and of $S$ on $Y$ are quantified, respectively, by $\eta$ and $\lambda$. The outcome variable is also influenced by the unmeasured psychosomatic state of the participant, represented by $P$. We allow $P$ to influence $Y$ via a direct path, quantified by $\tau$, and by an indirect path, mediated by $S$, and quantified by the product $\delta \, \lambda$. The combined effect of the direct and indirect paths represents the placebo effect. The direct path from $P$ to $Y$ represents the influence of the psychosomatic state on the outcome mediated by biochemical and physiological pathways distinct from the pathways influenced by the active treatment, while the influence of $P$ on $S$ allows for the possibility that $P$ also influences the same pathways targeted by the treatment $X$. (Experimental evidence that placebo effects influence biochemical pathways is provided, for example, in studies of placebo analgesia involving endogenous opioid systems\cite{price2008,levine1978,benedetti1996,grevert1983,levine1984,amanzio2001,benedetti1995}. See also figure 2 in reference\cite{finniss2010}, for empirical support about pathways influenced by both psychosocial context and drug treatments.)

The role played by the expectation-desire model of emotions is made explicit by the observed variables $E$, $D$, $I$ and $M$, representing, respectively, the expected symptom intensity, the desire for symptom improvement, the interaction between expectation and desire, and the emotional level (measured, for example, by the participant's mood). According to the expectation-desire model, $M$ is directly influenced by $E$, $D$, and their interaction $I = E \times D$. The causal influence of $M$ on $P$ is quantified by $\phi$.

In unblinded trials it is reasonable to expect that the treatment actually received by the participant will affect its expected symptom intensity, since participants who know they are receiving the active treatment will more likely experience an increase in their expectation to feel better. Hence, we include an arrow from $X$ to $E$. The implication is that the treatment can influence the outcome not only via the participant's somatic state, but also by its psychosomatic state via the paths $X \rightarrow E \rightarrow M \rightarrow P$ and $X \rightarrow E \rightarrow I \rightarrow M \rightarrow P$. The binary variable $Q$ represents the randomized psychological encouragement IV assuming the value 1 when a encouragement message (aiming to increase the desire for symptom improvement) is applied to the participant, and 0 otherwise.

In addition to the key variables described so far, it is important to recognize the existence of unmeasured confounders. Except for the exogenous variables $Z$ and $Q$, that by construction are not associated with any unmeasured confounders, the model includes confounders influencing all pairs of endogenous variables other than $I$, namely, $X$, $E$, $D$, $M$, $P$, $S$, and $Y$. (It is not necessary to include confounders between $I$ and the other endogenous variables, since $I$ is deterministically computed as the product of $E$ and $D$). For instance, $\bfU$ represents a set of unmeasured confounder variables influencing $X$ and $Y$. In order to avoid cluttering the figure, the confounder variables influencing $S$ and $P$ and all other endogenous variables are represented by the vector of variables $\bfmH = (\bfH_1, \ldots, \bfH_{11})^T$. (For the same reason the figure does not explicitly shows the error terms, which account for unmeasured variables influencing each particular variable in the model and are uncorrelated with each other). It would be unrealistic to assume, for example, that the emotion of a participant is determined by $E$, $D$, and $I$ alone. Hence, the model allows sets of unmeasured confounders, such as $\bfL_1$, $\bfL_2$ and $\bfL_3$, to influence emotion and expectation, emotion and desire, and expectation and desire, respectively. Similarly, it would be unrealistic to assume that emotion alone influences the psychosomatic state of a participant, and the model accommodates unmeasured confounders influencing these variables as well. Although, in practice, not all endogenous variables (other than $I$) will necessarily be influenced by confounders, the model still includes confounders for all 21 pairwise combinations of endogenous variables, since we want to derive estimators for the placebo and treatment effects under the most general setting possible.

In practice, however, it is impossible to accurately measure the unobserved somatic and psychosomatic states of a participant. Hence, Figure \ref{fig:dags}b presents a reduced version where $S$ and $P$ are not explicitly represented in the graph. Assuming linear relationships between $S$ and $X$, $P$ and $M$, and $Y$, $S$, and $P$, the causal influence of $X$ on $Y$ is given by $\beta = \eta \, \lambda$, while the influence of $M$ on $Y$ is given by $\psi = \phi \, \tau + \phi \, \delta \, \lambda$. Under this reduced model the instrumental variable $Q$ allows for the consistent estimation of the net placebo effect, $\psi$, using the IV estimator $\widehat{\psi} = \widehat{\cov}(Q, Y)/\widehat{\cov}(Q, M)$. Once the net placebo effect is estimated, it is possible to estimate the causal effect of $X$ on $Y$ using the IV estimator of the causal effect of $X$ on the residuals of the outcome variable after the removal of the placebo effect, $\widehat{\beta} = \widehat{\cov}(Z, \widehat{R})/\widehat{\cov}(Z, X)$, where $\widehat{R} = Y - \hat{\psi} \, M$ (see Methods for details).

\section{Performance evaluation}

\begin{figure}[!h]
$$
{\scriptsize \xymatrix@-1.4pc{
*+[F-:<10pt>]{Z} \ar[dr] &&& U \ar@{.>}[dll] \ar@{.>}[drr] && (a) &  & *+[F-:<10pt>]{Z} \ar[dr] &&& U \ar@{.>}[dll] \ar@{.>}[drr] && (b) &   \\
& *+[F-:<10pt>]{X} \ar[rrrr]^{\beta}  && && *+[F-:<10pt>]{Y} &  & & *+[F-:<10pt>]{X} \ar[rrrr]^{\beta} \ar[dddddd] && && *+[F-:<10pt>]{Y} &  \\
&&&&&&  & &&&&&&  \\
C_1 \ar@{.>}[uur] \ar@{.>}[ddddr] &&  &  &&& V_1 \ar@{.>}[uul] \ar@{.>}[ddddl]  & C_1 \ar@{.>}[uur] \ar@{.>}[ddddr] &&  &  &&& V_1 \ar@{.>}[uul] \ar@{.>}[ddddl]  \\
&& C_3 \ar@{.>}[dr] \ar@{.>}[uuul] && V_3 \ar@{.>}[dl] \ar@{.>}[uuur]  &&   & && C_3 \ar@{.>}[dr] \ar@{.>}[uuul] && V_3 \ar@{.>}[dl] \ar@{.>}[uuur]  &&   \\
C_2 \ar@{.>}[ddrrrrr] \ar@{.>}[uuuur] && L_1 \ar@{.>}[ddl] \ar@{.>}[r] & *+[F-:<10pt>]{M} \ar@/^1.25pc/[uuuurr]_{\psi} & L_2 \ar@{.>}[ddr] \ar@{.>}[l] &&  V_2 \ar@{.>}[uuuul] \ar@{.>}[ddlllll] & C_2 \ar@{.>}[ddrrrrr] \ar@{.>}[uuuur] && L_1 \ar@{.>}[ddl] \ar@{.>}[r] & *+[F-:<10pt>]{M} \ar@/^1.25pc/[uuuurr]_{\psi} & L_2 \ar@{.>}[ddr] \ar@{.>}[l] &&  V_2 \ar@{.>}[uuuul] \ar@{.>}[ddlllll] &  \\
&&&&&&  & &&&&&&  \\
& *+[F-:<10pt>]{E} \ar[rruu] \ar[rr] && *+[F-:<10pt>]{I} \ar[uu] && *+[F-:<10pt>]{D} \ar[lluu] \ar[ll] &   & & *+[F-:<10pt>]{E} \ar[rruu] \ar[rr] && *+[F-:<10pt>]{I} \ar[uu] && *+[F-:<10pt>]{D} \ar[lluu] \ar[ll] &   \\
&&& L_3 \ar@{.>}[ull] \ar@{.>}[urr] &&& *+[F-:<10pt>]{Q} \ar[lu]   & &&& L_3 \ar@{.>}[ull] \ar@{.>}[urr] &&& *+[F-:<10pt>]{Q} \ar[lu]   \\
&&&&&&  & &&&&&&  \\
*+[F-:<10pt>]{Z} \ar[dr] &&&  && (c) &  & *+[F-:<10pt>]{Z} \ar[dr] &&&  && (d) &   \\
& *+[F-:<10pt>]{X} \ar[rrrr]^{\beta}  && && *+[F-:<10pt>]{Y} &  & & *+[F-:<10pt>]{X} \ar[rrrr]^{\beta} \ar[dddddd] && && *+[F-:<10pt>]{Y} &  \\
&&&&&&  & &&&&&&  \\
  &&  &  &&&    &   &&  &  &&&    \\
&&   &&    &&   & &&   &&    &&   \\
  &&   & *+[F-:<10pt>]{M} \ar@/^1.25pc/[uuuurr]_{\psi} &  &&    &  &&   & *+[F-:<10pt>]{M} \ar@/^1.25pc/[uuuurr]_{\psi} &   &&    &  \\
&&&&&&  & &&&&&&  \\
& *+[F-:<10pt>]{E} \ar[rruu] \ar[rr] && *+[F-:<10pt>]{I} \ar[uu] && *+[F-:<10pt>]{D} \ar[lluu] \ar[ll] &   & & *+[F-:<10pt>]{E} \ar[rruu] \ar[rr] && *+[F-:<10pt>]{I} \ar[uu] && *+[F-:<10pt>]{D} \ar[lluu] \ar[ll] &   \\
&&&   &&& *+[F-:<10pt>]{Q} \ar[lu]   & &&&   &&& *+[F-:<10pt>]{Q} \ar[lu]   \\
}}
$$
\caption{Models used in the simulation study. Panels a and b represent, respectively, blinded and unblinded trials influenced by confounders. For simplicity we include a single confounder variable per pair of endogenous variables (other than $I$), but still simulate confounding across the 10 possible pairwise combinations of the endogenous variables $X$, $Y$, $E$, $M$, and $D$. Panels c and d represent, respectively, unconfounded blinded and unblinded trials. For simulations under the null $H_0: \psi = 0$ there are no arrows from $M$ to $Y$. Similarly, for  simulations under $H_0: \beta = 0$, there are no arrows from $X$ to $Y$.}
\label{fig:simdags}
\end{figure}
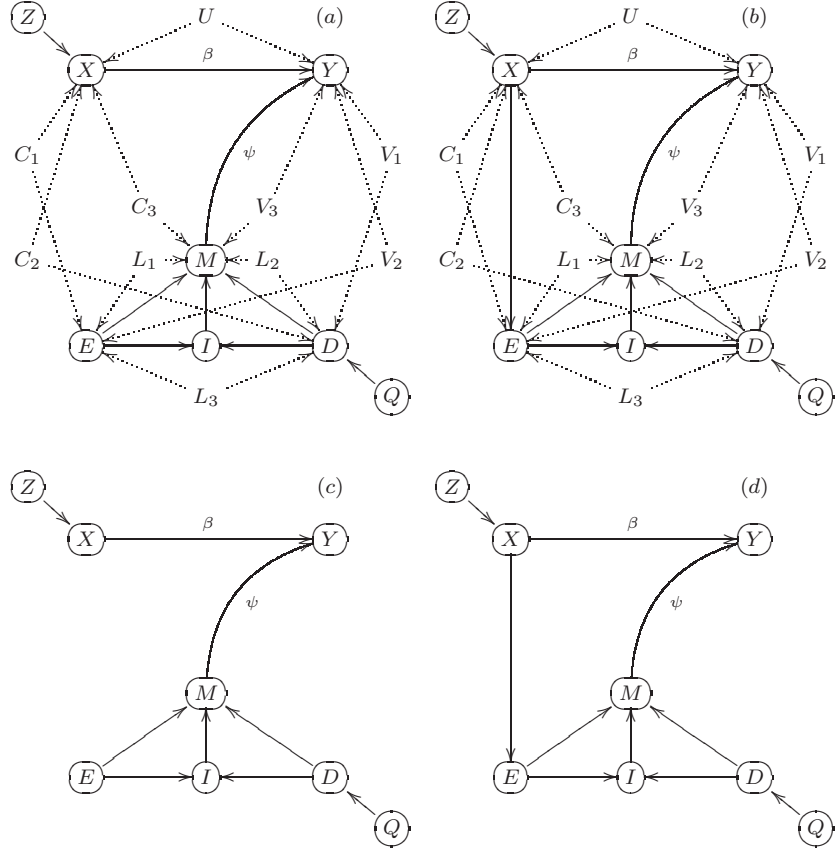

We assessed the statistical performance of the proposed method (and compare it to a naive regression approach) in 16 simulation experiments evaluating the empirical type I error rate and empirical power of randomization tests for the null hypotheses that the placebo effect is zero, $H_0: \psi = 0$, and that the treatment effect is zero, $H_0: \beta = 0$. Descriptions of the randomization tests and simulation experiments are provided in the Methods. We simulated data from blinded and unblinded trials, in the presence and absence of confounders, according to the models presented in Figure \ref{fig:simdags}.

For each setting, we ran 4 separate simulation experiments generating data: (i) under the null for treatment and placebo effects; (ii) under the alternative for treatment, and null for placebo effects; (iii) the other way around; and (iv) under the alternative for treatment and placebo effects. Each simulation experiment employed 10,000 distinct synthetic data sets with diverse characteristics (see Methods). Although the randomization tests are non-parametric procedures free of distributional assumptions, we still generated data using gaussian errors in order to met the distributional requirements of the regression based analytical tests used in our comparisons.

\begin{figure}[!h]
\begin{center}
\includegraphics[angle=270, scale = 0.61, clip]{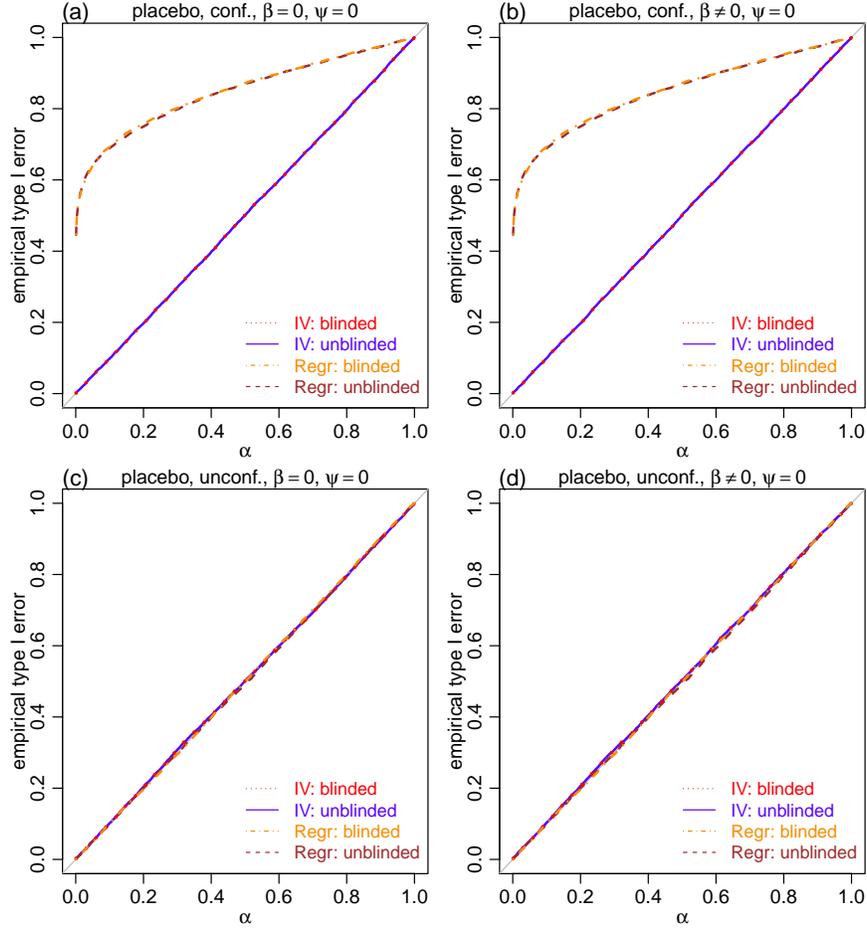}
\caption{Empirical type I error rates of the placebo effect null, $H_0: \psi = 0$, in both blinded and unblinded settings. Panels a and b show that, in the presence of confounders, the type I error rate of the IV approach is controlled at the exact nominal level (red and blue), whereas the regression based test leads to highly inflated error rates (orange and brown). Panels c and d show that, in the absence of confounding, both IV and regression approaches show well controlled errors. The nominal significance level is represented by $\alpha$.}
\label{fig:simpsi}
\end{center}
\end{figure}

\begin{figure}[!h]
\begin{center}
\includegraphics[angle=270, scale = 0.61, clip]{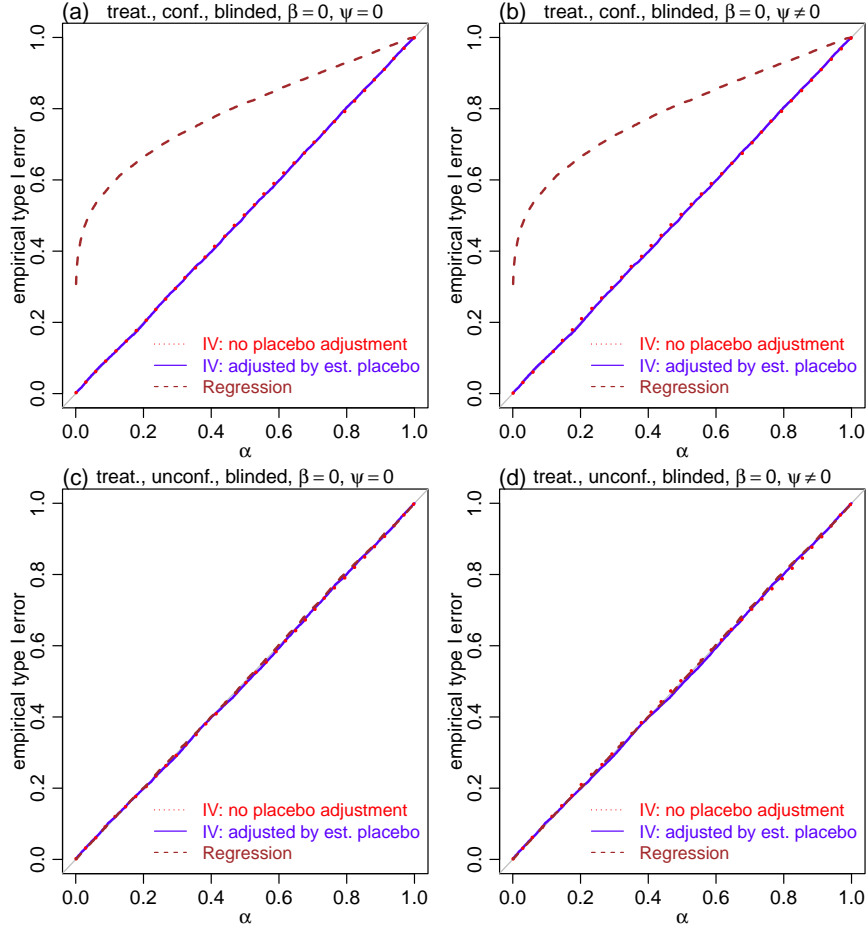}
\caption{Empirical type I error rates for the treatment effect null, $H_0: \beta = 0$, in the blinded setting. Panels a and b show that, in the presence of confounders, the type I error rates of the IV approaches are controlled at the exact nominal level (red and blue), whereas the regression based test leads to highly inflated error rates (brown). Panels c and d show that, in the absence of confounding, both IV and regression approaches show well controlled errors. The nominal significance level is represented by $\alpha$.}
\label{fig:simbetablinded}
\end{center}
\end{figure}

\begin{figure}[!h]
\begin{center}
\includegraphics[angle=270, scale = 0.61, clip]{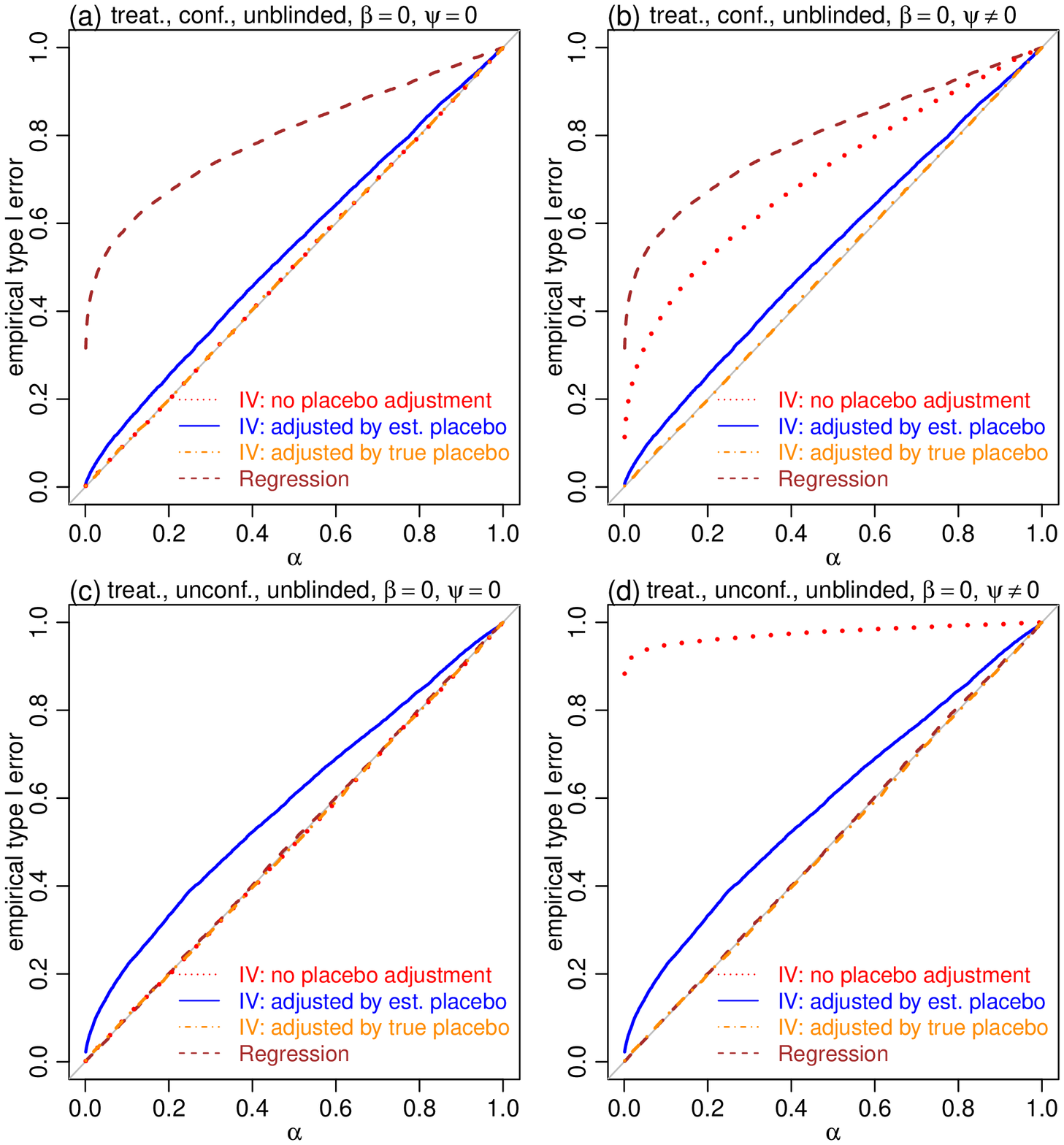}
\caption{Empirical type I error rates for the treatment effect null, $H_0: \beta = 0$, in the unblinded setting.  The two-step IV approach (blue) shows slightly inflated errors in the presence (panels a and b) and absence (panels c and d) of confounders. Note that the larger errors in panels c and d, in comparison to a and b, are likely due to the effective stronger influence of $X$ on $M$ in the simulations unaffected by confounders (the presence of confounders can considerably increase the amount of noise), so that adjustment by $\hat{\psi}$ leaks more information about $X$ in the absence than in the presence of confounders. The estimator adjusted by the true placebo effect (dark-orange) leads, nonetheless, to well controlled errors. The non-adjusted IV approach (red) leads to well controlled errors in the absence of placebo effects (panels a and c), but to highly inflated errors in the presence of placebo effects (panels b and d). Regression (brown) leads to highly inflated errors in the presence of confounders (panels a and b), but to well controlled error rates in their absence (panels c and d).}
\label{fig:simbetaunblinded}
\end{center}
\end{figure}

Figure \ref{fig:simpsi} presents the results for the placebo effect tests, and shows that the error rates of the IV approach (red and blue) are controlled at the exact nominal levels in both blinded and unblinded settings, in the presence and absence of confounders. The regression approach (brown and dark-orange), on the other hand, shows highly inflated errors in the presence of confounders (panels a and b), since the association between $M$ and $Y$, caused by confounders, is mistaken by an influence of $M$ on $Y$. Being able to control type I error rates at the exact nominal level is a desirable statistical property, as it means that the test is neither conservative nor liberal.

Figure \ref{fig:simbetablinded} presents the results for the treatment effect tests in the blinded setting. In addition to the two-step estimator (blue), we also evaluated the simple IV estimator $\widehat{\beta} = \widehat{\cov}(Z, Y)/\widehat{\cov}(Z, X)$, which does not account for the placebo effect (red). The results show, again, well controlled error rates for both IV approaches, but inflated errors for the regression test (brown) in the presence of confounders (panels a and b).

Figure \ref{fig:simbetaunblinded} presents the results for the unblinded case. All panels show slightly inflated errors for the two-step IV estimator (blue). The likely reason is that the estimated placebo effects are noisy and unable to completely block the influence of $X$ on $Y$ through the paths mediated by $M$. To test this supposition, we evaluated an additional IV estimator, where the true placebo effect was used in the computation of the residuals (i.e., we estimated $\beta$ by $\widehat{\beta} = \widehat{\cov}(Z, R)/\widehat{\cov}(Z, X)$, where $R = Y - \psi \, M$, instead of $\widehat{\beta} = \widehat{\cov}(Z, \widehat{R})/\widehat{\cov}(Z, X)$, where $\widehat{R} = Y - \hat{\psi} \, M$). Results based on this estimator (dark-orange) show that, indeed, adjustment by the true placebo effect leads to error rates controlled at the nominal level. The regression approach (brown) shows again highly inflated errors in the presence of confounders (panels a and b). Panels a and c show well controlled errors for the non-adjusted IV estimator (red) in the absence of placebo effects as, in this case, there are no paths from $X$ to $Y$, and the association between $X$ and $Y$ induced by confounders is accounted by the IV estimator. Panels b and d, on the other hand, show highly inflated error rates in the presence of placebo effects since, in this case, $X$ can influence $Y$ through the paths mediated by $M$.

These observations suggest that, in practice, when analyzing the results of unblinded trials, we should first test for the existence of placebo effect, and then use the two-step IV estimator if $H_0: \psi = 0$ is rejected, and the non-adjusted one if $H_0: \psi = 0$ is accepted. While this strategy can decrease the chance of the two-step approach making a type I error in the absence of placebo effects, the estimator is still unable to avoid slightly inflated errors produced in the presence of placebo effects. We point out, however, that the two-step procedure still represents a strong improvement over the alternative approach of not adjusting for placebo effects in the presence of confounders (compare the red and blue curves in panel b of Figure \ref{fig:simbetaunblinded}).

\begin{figure}[!b]
\begin{center}
\includegraphics[angle=270, scale = 0.61, clip]{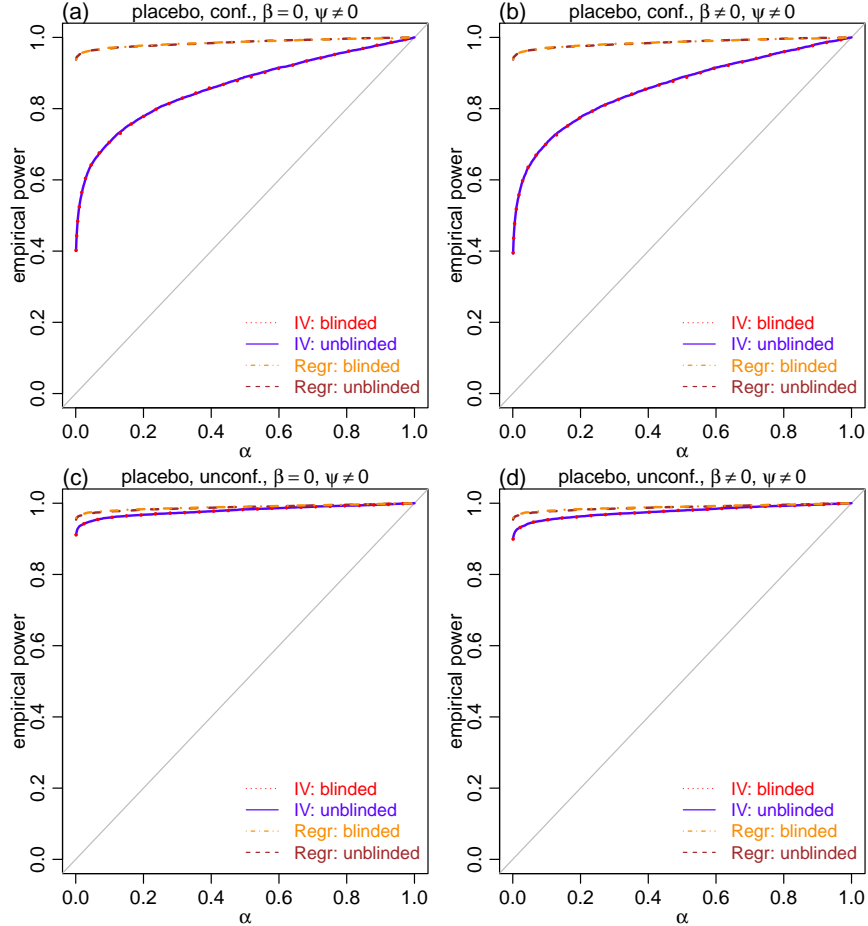}
\caption{Empirical power to detect placebo effects in the blinded and unblinded settings. Panels a and b show the results in the presence of confounders, whereas panels c and d show the results in their absence. The regression approach (brown and dark-orange) were considerably better powered than the IV approaches (blue and red) in the presence of confounders (panels a and b), but only slightly better powered in the absence of confounders (panels c and d). Both regression and IV approaches showed similar power under the blinded and unblinded settings.}
\label{fig:powerpsi}
\end{center}
\end{figure}

\begin{figure}[!h]
\begin{center}
\includegraphics[angle=270, scale = 0.61, clip]{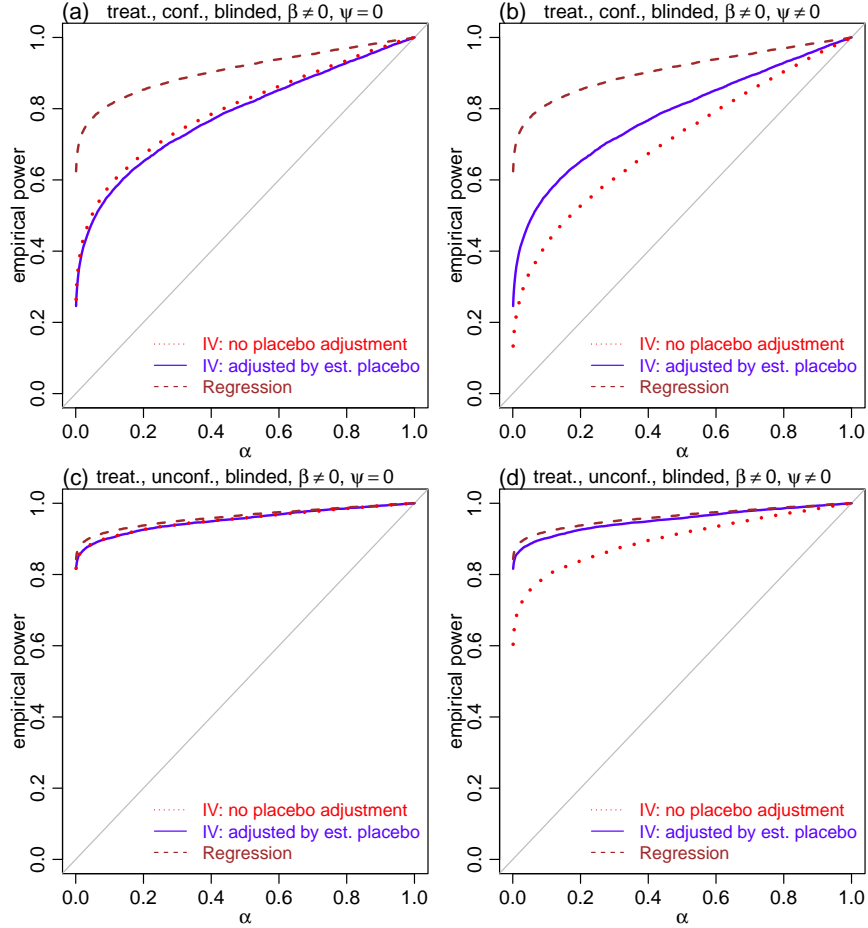}
\caption{Empirical power for detecting treatment effects in the blinded setting. The regression approach (brown) tends to be better powered than the IV approaches in the presence of confounders (panels a and b), but only slightly better in the absence of confounding (panels c and d). The two-step IV approach (blue) tends to be better powered than the non-adjusted one (red) in the presence of placebo effects (panels b and d), but both IV approaches tend to be comparable in absence of placebo effects (panels a and c).}
\label{fig:powerbetablinded}
\end{center}
\end{figure}

\begin{figure}[!h]
\begin{center}
\includegraphics[angle=270, scale = 0.61, clip]{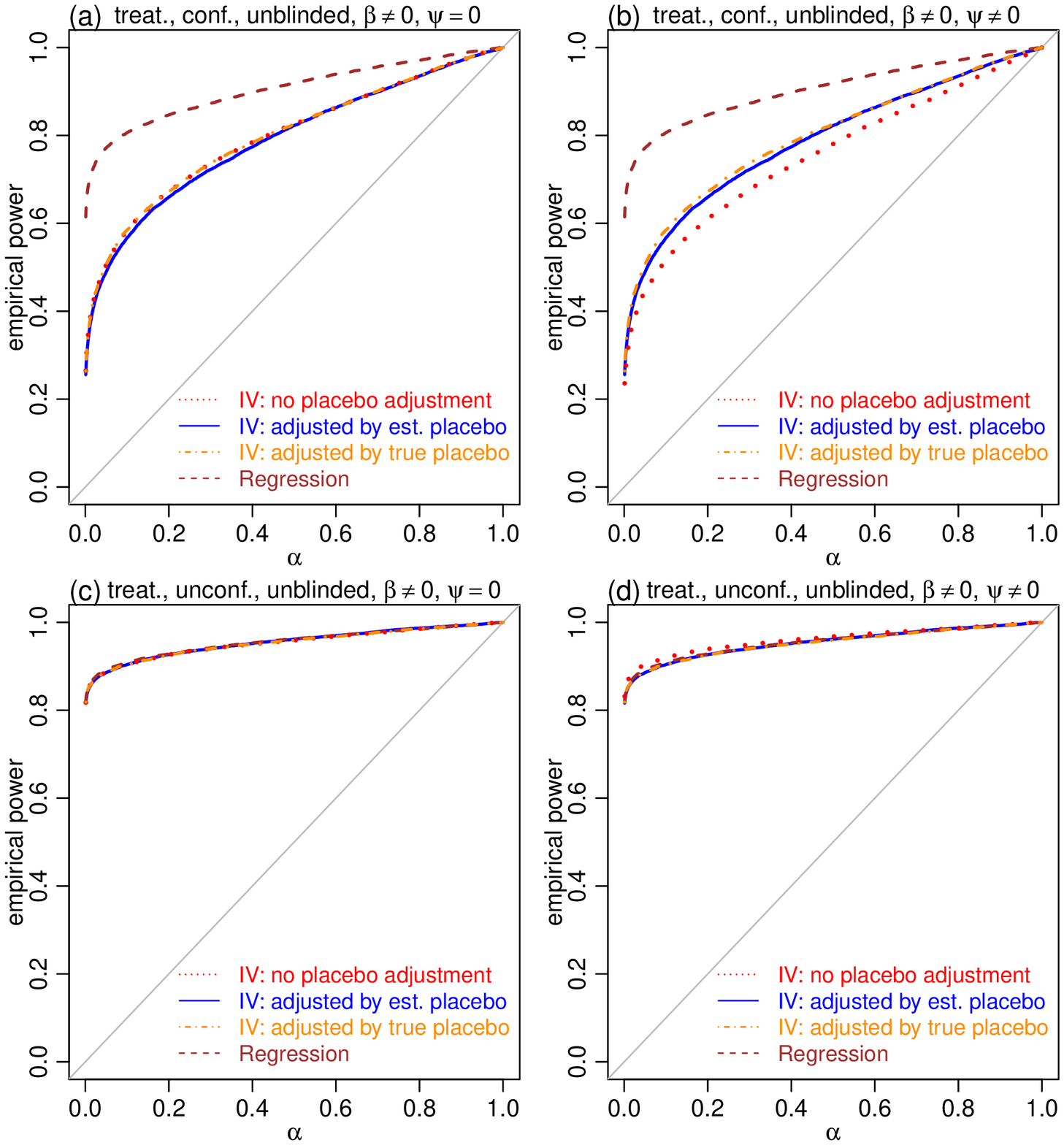}
\caption{Empirical power for detecting treatment effects in the unblinded setting. The regression approach (brown) tended to be better powered than the IV approaches in the presence of confounders (panels a and b), but comparable in the absence of confounding (panels c and d). The two-step IV approach (blue) tended to be slightly better powered than the non-adjusted one (red) in the presence of placebo effect (panel b), but comparable in the other panels.}
\label{fig:powerbetaunblinded}
\end{center}
\end{figure}

\begin{figure}[!h]
\begin{center}
\includegraphics[angle=270, scale = 0.61, clip]{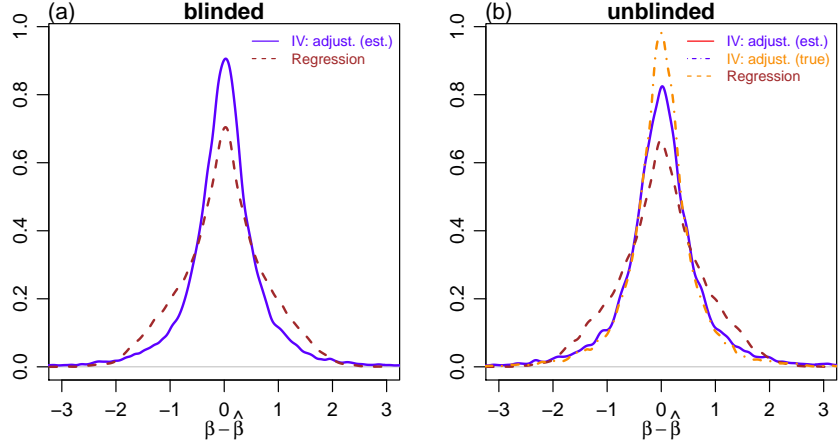}
\caption{Comparison of the bias of the regression and IV estimators. Panels a and b show the densities of the difference between true and estimated treatment effects, $\beta - \hat{\beta}$, in the blinded and unblinded settings, respectively. In both settings we observed larger bias in the regression estimates, in comparison to the IV approaches, as illustrated by the heavier tails of the brown densities.}
\label{fig:deltabeta}
\end{center}
\end{figure}

For completeness, we also report an evaluation of the empirical power (Figures \ref{fig:powerpsi}, \ref{fig:powerbetablinded}, and \ref{fig:powerbetaunblinded}). We point out, however, that power results are more sensitive to the choice of parameter values employed in the generation of the simulated data (e.g., sample size, the strength of treatment, placebo and confounding effects, and etc), than the type I error rates. In any case, these empirical power results, still serve to illustrate some general patterns. For instance, the regression tests tended to show considerably stronger power than the IV approaches in the presence of confounders (compare the brown and blue curves in panels a and b of Figures \ref{fig:powerbetablinded} and \ref{fig:powerbetaunblinded}). We point out, however, that this increased power is likely an artifact of the biased estimates of $\beta$ outputted by the regression approach. Figure \ref{fig:deltabeta}, illustrates how the regression estimates tended to show larger bias than the estimates generated by the IV estimators (note the heavier tails of the brown density, in both blinded and unblinded cases). In other words, the increased power is likely a consequence of the overestimation of the treatment effect by the regression approach, which mistakenly interprets the association between treatment and outcome caused by unmeasured confounders as a stronger influence of the treatment on the outcome.


\begin{figure}[!h]
\begin{center}
\includegraphics[angle=270, scale = 0.61, clip]{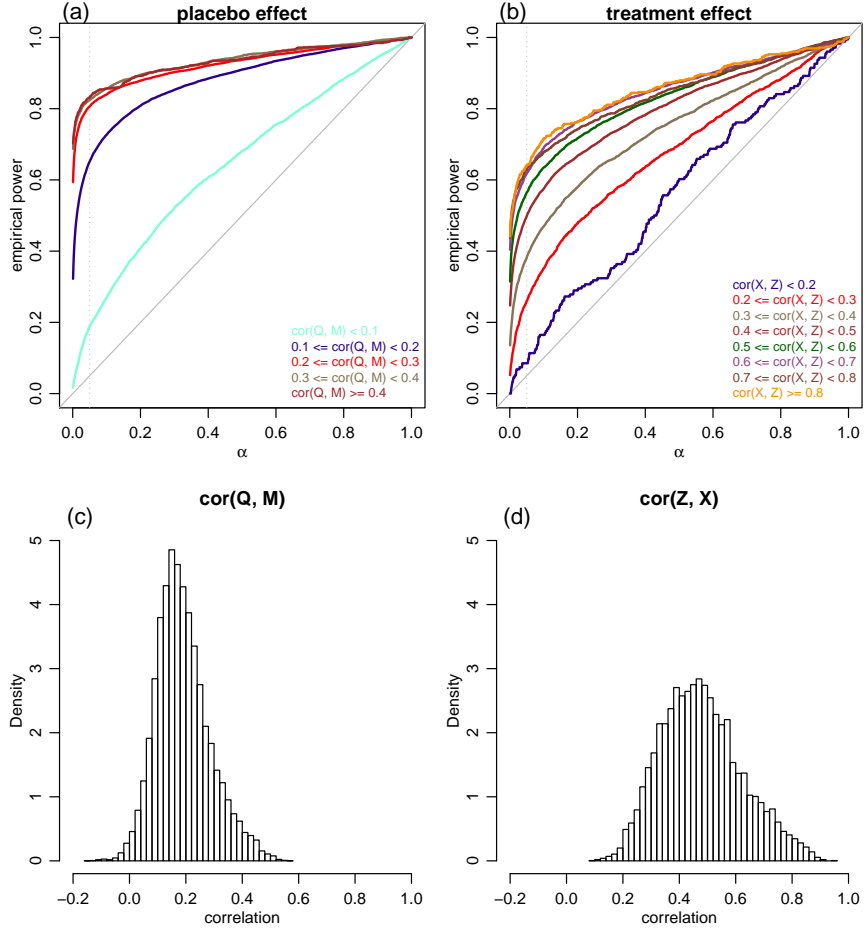}
\caption{Empirical power curves stratified by strength of association with the IV variable. Panel a shows the power curves for the placebo effect IV estimator $\hat{\psi}$, stratified according to the correlation between $Q$ and $M$ (panel c shows the distribution of the correlation between $Q$ and $M$ across all simulations used to construct the power curves in panel a). Panel b shows the power curves for the two-step treatment effect IV estimator $\hat{\beta}$, stratified according to the correlation between $Z$ and $X$ (panel d shows the distribution of the correlation between $Z$ and $X$ over the simulations used to estimate the power curves in panel b). Results based on blinded and unblinded simulations influence by confounders.}
\label{fig:stratifiedpower}
\end{center}
\end{figure}

At least for the parameter ranges adopted in our simulations, we observed good empirical power of the IV approach to detect placebo effects, even when the correlation between psychological encouragement and emotional level was relatively low (Figure \ref{fig:stratifiedpower}a). This suggests that the psychological encouragement treatment does not need to be highly effective in manipulating the emotional levels, in order for the approach to work well in practice. Similarly, Figure \ref{fig:stratifiedpower}b shows good empirical power of the two-step IV approach to detect treatment effects when the correlation between the assigned and received treatment is moderate, suggesting that the proposed approach does not require high levels of compliance in order to perform well.

\begin{figure}[!h]
\includegraphics[angle=270, scale = 0.61, clip]{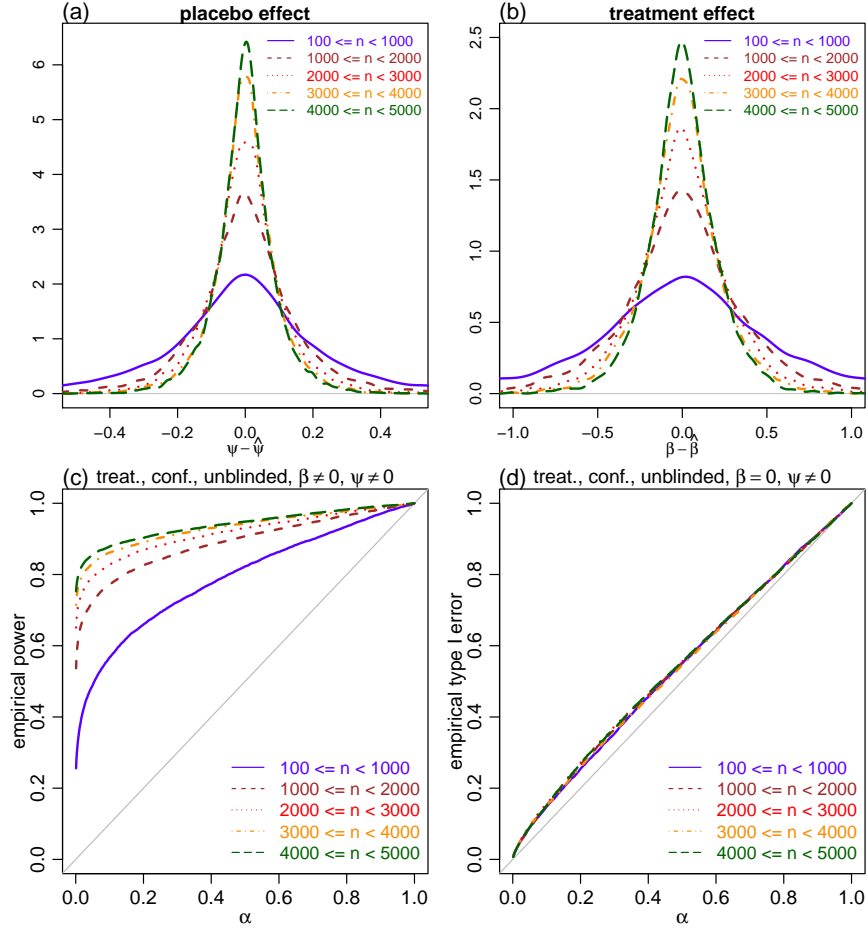}
\caption{Consistency of the $\hat{\psi}$ and $\hat{\beta}$ estimators. Panels a and b present, respectively, the densities of $\psi - \hat{\psi}$ and $\beta - \hat{\beta}$ for 5 increasing sample size ranges, and illustrate the consistency of the $\hat{\psi}$ and $\hat{\beta}$ estimators (which tend to get closer to the true parameter values as the sample size increases). Panel c shows that, as expected, the statistical power to detect a treatment effect increases with the sample size. Panel d, on the other hand, shows that increasing sample sizes do not reduce type I error rates, even though we are able to better estimate the placebo effects. The likely reason is that while larger sample sizes lead to better $\hat{\psi}$ estimates, they also increase the statistical power to detect very small effects, so that the advantage of a more precise estimate of $\hat{\psi}$ is counterbalanced by the increased propensity to detect small and spurious treatment effects as true signals. Results were based on data simulated from unblinded trials influenced by placebo effects and counfounders, as described in the Methods section.}
\label{sfig:suppleconsitency}
\end{figure}

\begin{figure}[!b]
\includegraphics[angle=270, scale = 0.61, clip]{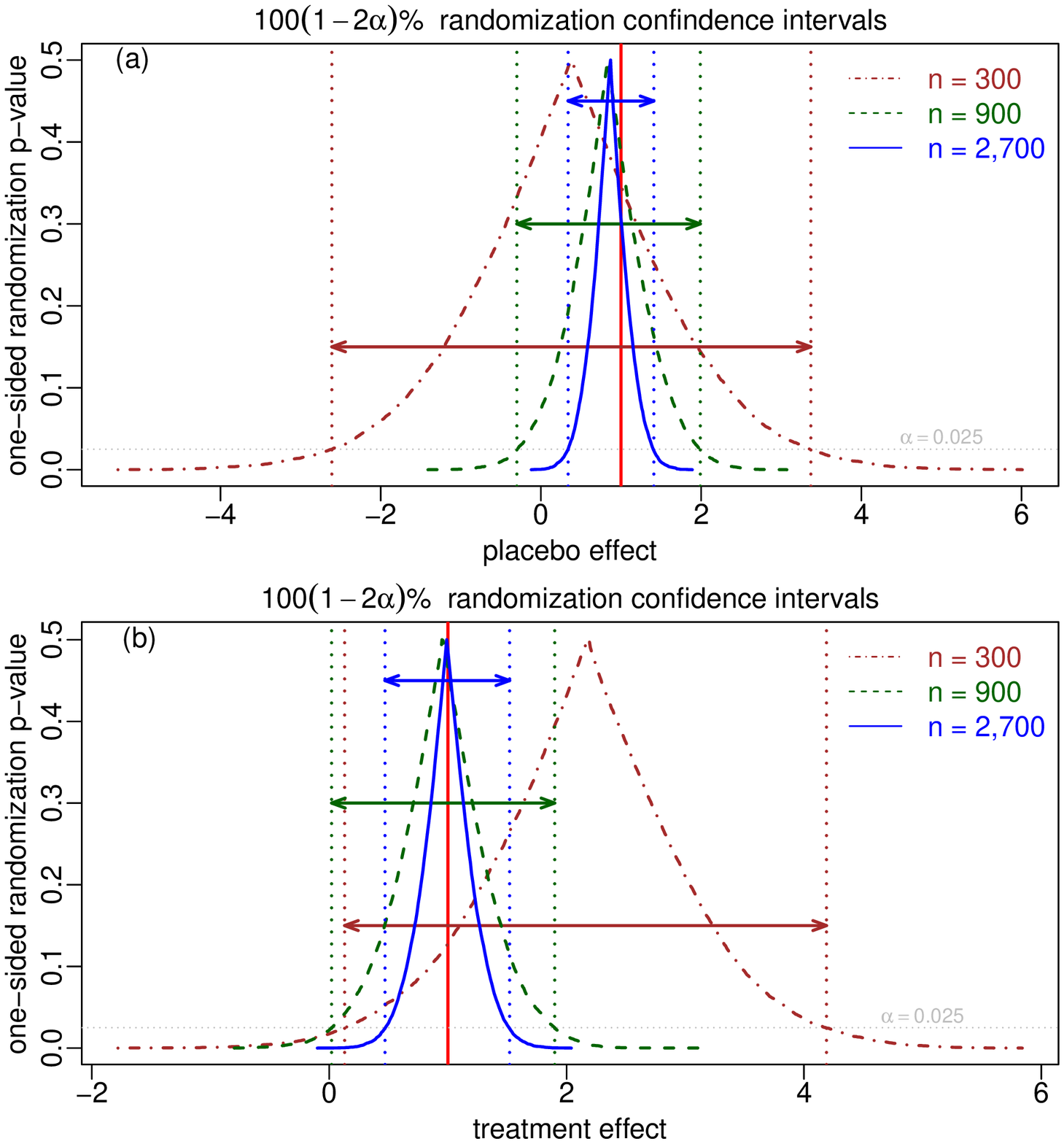}
\caption{Randomization confidence intervals for placebo and treatment effects. The brown, dark-green and blue curves show the one-sided p-value profiles derived from randomization tests for 3 simulated data sets of increasing sizes (300, 900, and 2,700, respectively), generated under the unblinded setting influenced by confounders (all simulation parameters, other than sample size, were set to 1). The 95\% confidence intervals for the placebo (panel a) and treatment effects (panel b) are shown by the respective double-headed colored arrows. The red vertical line corresponds to the true parameter values, $\psi = 1$ and $\beta = 1$.}
\label{fig:randconfint}
\end{figure}

A natural question, at this point, is whether larger sample sizes (and, hence, more precise estimates of $\hat{\psi}$) would be able to decrease the slightly inflated error rates produced by the two-step estimator in unblinded trials. Figure \ref{sfig:suppleconsitency} presents additional simulation experiments showing that, while the empirical power and the $\hat{\psi}$ and $\hat{\beta}$ estimates are greatly improved by larger sample sizes, the type I error rates stay roughly the same (likely because larger sample sizes increase the ability of a test to detect small effects, since the randomization null distributions tend to be more concentrated around 0, so that the improved $\hat{\psi}$ estimates are counterbalanced by the increased propensity to detect small and spurious treatment effects). These results suggest that special care must be taken while interpreting the results of hypothesis tests in the unblinded case, even for large sample sizes. In any case, when the goal is estimation rather than testing, the consistency of the two-step estimator guarantees that the treatment estimates will converge to the true value as the sample size increases.

This observation is particularly important in view of the current trend in the biomedical field, where researchers are shifting from relying exclusively in p-values and are paying more attention to parameter estimates and confidence intervals. To meet this latter need, we also describe in the Methods how to generate confidence intervals (CIs) for placebo and treatment effects by inverting randomization tests. Figure \ref{fig:randconfint} shows 95\% CIs for the placebo and treatment effects, from 3 simulated data sets of increasing sizes. The randomization CIs inherit the statistical properties of the randomization tests, hence, the placebo effect CIs (and treatment effect CIs from blinded trials) are exact in the sense that a $100 (1 - \alpha)$\% interval will contain the true parameter value $100 (1 - \alpha)$\% of the time. Note that while the treatment effect CIs from unblinded trails won't be exact, they are still going to be centered around the estimated treatment effect, which will, nevertheless, converge to the true value as the sample size increases.

\section{Discussion}

Clinical trials traditionally employ blinding to control the influence of placebo effects. It has being pointed out, however, that even blinded studies might be influenced by placebo effects, as the patients perceptions and beliefs about the treatment they think they received are able to trigger strong placebo effects\cite{price2008,mcrae2004,bausell2005}. Recently, a number of statistical approaches have been proposed to quantify the contributions of treatment and placebo effects to a clinical outcome\cite{zhang2013,jamshidian2014,liu2016}. These approaches, nonetheless, are tailored to blinded trials, and leverage blinding assessment data to quantify the amount of unmasking taking place during the trial. Our IV approach, on the other hand, allows the quantification of treatment and placebo effects not only in blinded, but also in unblinded trials.

The key idea underlying the IV approach (what actually allows the consistent estimation of both treatment and placebo effects in the presence of unmeasured confounders), is the use of randomization to separately manipulate the treatment assignment and encouragement messages. In this sense, the proposed approach is similar in spirit (but not exactly equivalent) to a randomized treatment-belief trial (RTB)\cite{roy2012}, where the treatment assignment is manipulated by randomization, whereas the belief is manipulated by varying the allocation ratio of participants assigned to control and treatment groups in a, necessarily, blinded trial. Hence, our IV approach can be viewed as a more flexible type of RTB that is applicable to both blinded and unblinded studies, and might be easier to administer than a standard RTB, which requires the stratification of study participants over several arms with distinct treatment/control allocation ratios in order to be able to assess placebo effects.

The proposed IV approach enjoys appealing statistical properties. The IV estimators are consistent, meaning that the estimates converge to the true values as sample size increases. The randomization tests for placebo effects are exact in both blinded and unblinded trails, whereas the treatment effect tests are exact in blinded trials, but slightly liberal in unblinded ones. Furthermore, the confidence intervals obtained by inverting randomization tests inherit these appealing properties.

\begin{figure}[!h]
$$
{\tiny \xymatrix@-1.3pc{
*+[F-:<10pt>]{Z} \ar[dr] &&& U \ar@{.>}[dll] \ar@{.>}[drr] && (a) &  \\
& *+[F-:<10pt>]{X} \ar[rrrr]^{\beta} \ar[dddddd] \ar[rrd] && && *+[F-:<10pt>]{Y} &  \\
&& C_4 \ar@{.>}[r] \ar@{.>}[lu]  & *+[F-:<10pt>]{A} \ar[rru]^{\kappa} & V_4 \ar@{.>}[l] \ar@{.>}[ru] && \\
C_1 \ar@{.>}[uur] \ar@{.>}[ddddr] &&& *+[F-:<10pt>]{W} \ar[u] &&& V_1 \ar@{.>}[uul] \ar@{.>}[ddddl]  \\
&& C_3 \ar@{.>}[dr] \ar@{.>}[uuul] & & V_3 \ar@{.>}[dl] \ar@{.>}[uuur]  &&  \\
C_2 \ar@{.>}[ddrrrrr] \ar@{.>}[uuuur] && L_1 \ar@{.>}[ddl] \ar@{.>}[r] & *+[F-:<10pt>]{M} \ar@/^0.25pc/[uuuurr]_{\psi} & L_2 \ar@{.>}[ddr] \ar@{.>}[l] &&  V_2 \ar@{.>}[uuuul] \ar@{.>}[ddlllll] \\
&&&&&& \\
& *+[F-:<10pt>]{E} \ar[rruu] \ar[rr] && *+[F-:<10pt>]{I} \ar[uu] && *+[F-:<10pt>]{D} \ar[lluu] \ar[ll] & \\
&&& L_3 \ar@{.>}[ull] \ar@{.>}[urr] &&& *+[F-:<10pt>]{Q} \ar[lu] \\
}}
$$
\includegraphics[angle=270, scale = 0.61, clip]{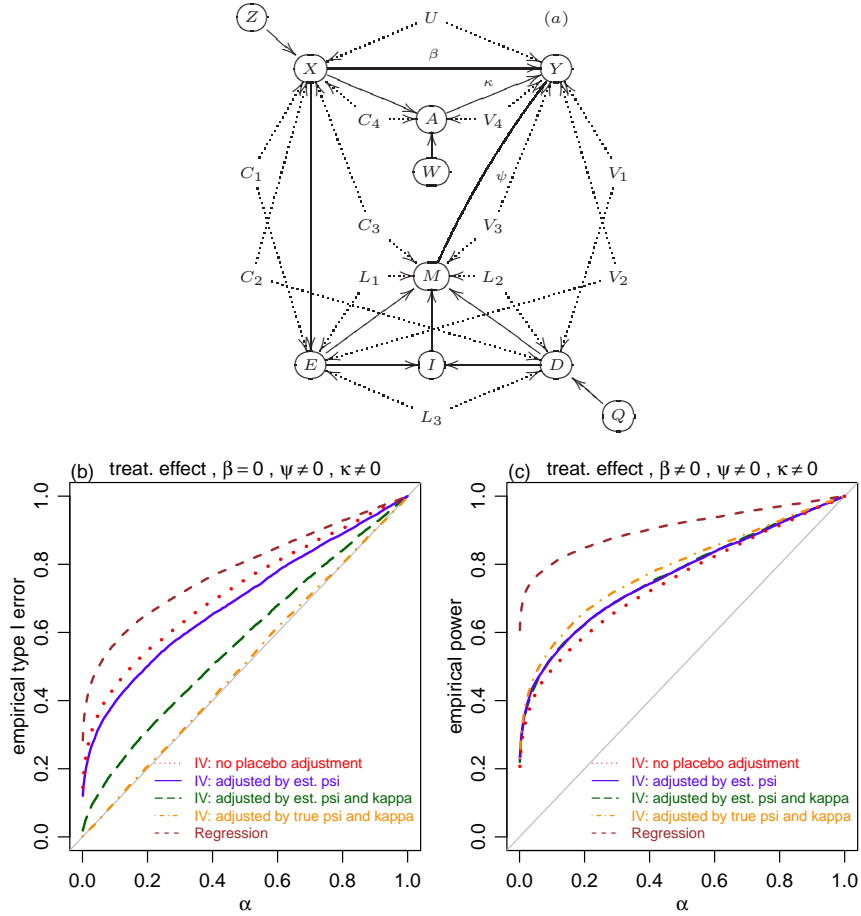}
\caption{A more complex example. Panel a presents a more complex model where the placebo effect is mediated by $M$ (according to the desire-expectation model) but also by an additional variable $A$. Assuming that a randomized instrument, $W$, is available to manipulate $A$, we can estimate the treatment effect using the estimator $\hat{\beta}^\ast = \widehat{\cov}(Z, \hat{R}^\ast)/\widehat{\cov}(Z, X)$ where $\hat{R}^\ast = Y - \hat{\psi} \, M - \hat{\kappa} \, A$. Panel b shows the empirical type I error rates for a simulation experiment under the unblinded setting influenced by confounders. The IV estimator adjusted by the true $\psi$ and $\kappa$ values is able to control error rates at the nominal levels (dark-orange). The IV estimator adjusted by $\hat{\psi}$ and $\hat{\kappa}$ shows slightly inflated errors (dark-green). As expected, adjustment with $\hat{\psi}$ alone (blue) leads to higher error rates than adjustment with both $\hat{\psi}$ and $\hat{\kappa}$. Similarly, the IV estimator using no adjustment (red) has higher errors than adjustment by $\hat{\psi}$ alone. The regression based estimator (brown) is adjusted by both $M$ and $A$ covariates, but still leads to inflated errors due to the presence of confounders. Panel c shows the empirical power results.}
\label{sfig:suppledag}
\end{figure}

An implicit assumption of the model in Figure \ref{fig:dags}a is that the placebo effect is mediated exclusively by the interplay of desire, expectation, and emotion. While it is believed that the desire-expectation model plays a key role in the triggering of placebo effects, other mechanisms, such as conditioning, might also be at work\cite{price2008,finniss2010}. Clearly, when this is the case, a treatment effect estimate, adjusted by the desire-expectation component alone, will still be biased (although less biased than the estimate computed without accounting for it). In any case, if we are also able to assess and measure these additional mechanisms, then the same statistical framework can be used to obtain consistent estimates of treatment effects in the presence of confounders (we only need additional IVs to manipulate the additional placebo related variables). Figure \ref{sfig:suppledag} shows an example.

From a pragmatic perspective, the proposed method is easy to implement. It only requires the ability to assess expectation, desire, and emotion, as well as, the development of a psychological encouragement IV, capable of manipulating the level of desire of a study participant. For example, in trials run into a clinic, a simple encouragement conversation with a caregiver would work as the ``active treatment" of the psychological encouragement IV. The desire and emotional level could then be recorded by a questionnaire or interview after the encouragement treatment, but prior to the measurement of the outcome variable.

Another application of the proposed method (the one that actually motivated this work) is in the personalized monitoring of treatment response in mobile health. The statistical validity of using treatment assignment as an IV, in the context of longitudinal data provided by a single patient, has been established in reference\cite{chaibubneto2016}. However, as pointed out by the authors, it is impossible to disentangle treatment and placebo effects based on the treatment assignment IV alone, since it is impossible to blind the patient to a self administered treatment. Implementation of the proposed IV approach in mobile health applications is also strait-forward. For instance, the psychological treatment could be delivered by encouragement messages popping up in the screen of a smartphone (according to a randomized schedule, where every day the participant has an equal chance of receiving, or not, the encouragement message), and the measurement of the emotional and desire levels can be assessed by short electronic surveys/quentionnaires delivered by the participant's smartphone on a daily basis. We expect the proposed method to play an important role in these personalized medicine\cite{schork2015,topol2012} applications.

Finally, for both (population-based) clinical trials and personalized monitoring of treatment response, the instrument $Q$ serves the double role of disentangling placebo from treatment effects, and increasing the desire for improved symptoms. This latter capacity can induce a placebo effect and ultimately lead to more positive clinical outcomes. While the manipulation of the expectation for symptom intensity could, in principle, be used to consistently estimate a placebo effect under the proposed approach (i.e., we could have an IV influencing $E$ instead of $D$), the manipulation of expectation levels needs to be accompanied by the honest disclosure of the expected benefits of a treatment (and, in some cases, might raise ethical issues)\cite{finniss2010}. Manipulation of the desire for improved symptoms, on the other hand, provides an ethically defensible practice in the design of clinical trials and in the personalized monitoring of patients.

\section{Methods}

\subsection{Identification of causal effects using instrumental variables}

We subscribe to the mechanism-based account of causation championed by Pearl\cite{pearl2000}. In this framework, the qualitative description of the assumptions regarding the causal relations between the variables involved in our proposed method, is encoded by the directed acyclic graph presented in Figure \ref{fig:dags}a. Assuming a linear relation between the outcome, $Y$, and the unobserved somatic and psychosomatic state variables, $S$ and $P$, we have that,
\begin{equation}
Y = \mu_Y + \lambda \, S + \tau \, P + f_Y(\bfU, \bfmV, \bfmH) + \epsilon_Y~,
\label{eq:linearYfull}
\end{equation}
where $\bfmV = (\bfV_1, \bfV_2, \bfV_3)^T$, $\bfmH = (\bfH_1, \ldots, \bfH_{11})^T$, $\epsilon_Y$ represents an error term accounting for the unmeasured variables influencing exclusively $Y$, and $f_Y(\bfU, \bfmV, \bfmH)$ represents is a general scalar function of the variables in $(\bfU, \bfmV, \bfmH)$ influencing $Y$.

Since $S$ and $P$ are unobserved variables, we need to derive the reduced model for the outcome variable that is not a function of $S$ and $P$. Assuming a linear relation between $P$ and $M$, and between $S$ and $P$ and $X$, we have that,
\begin{align}
P &= \mu_P + \phi \, M + f_P(\bfmH) + \epsilon_P~, \label{eq:linearP} \\
S &= \mu_S + \eta \, X + \delta \, P + f_S(\bfmH) + \epsilon_S~, \label{eq:linearS}
\end{align}
where $f_P(\bfmH)$ and $f_S(\bfmH)$ are arbitrary scalar functions of $\bfmH$, and $\epsilon_P$ and $\epsilon_S$ are the respective error terms influencing $P$ and $S$, respectively (we also assume that all variable specific error terms are uncorrelated).

Substituting equations (\ref{eq:linearP}) and (\ref{eq:linearS}) into equation (\ref{eq:linearYfull}), we obtain the reduced outcome model,
\begin{equation}
Y = \mu_Y^\ast + \beta \, X + \psi \, M + f^\ast(\bfU, \bfmV, \bfmH) + \epsilon_Y^\ast
\label{eq:linearYreduced}
\end{equation}
where $\beta = \eta \, \lambda$, $\psi = \phi \, \tau + \phi \, \delta \, \lambda$, $\mu_Y^\ast = \mu_Y + \lambda \, \mu_S + (\tau + \delta \, \lambda) \, \mu_P$, $\epsilon_Y^\ast = \epsilon_Y + \lambda \, \epsilon_S + (\tau + \delta \, \lambda) \, \epsilon_P$, and $f^\ast(\bfU, \bfmV, \bfmH) = f_Y(\bfU, \bfmV, \bfmH) + \lambda \, f_S(\bfmH) + (\tau + \delta \, \lambda) \, f_P(\bfmH)$. Equation (\ref{eq:linearYreduced}) represents the outcome model in Figure \ref{fig:dags}b.

Because the instrumental variable $Q$ is randomized, and hence statistically independent of any variables that are not directly or indirectly influenced by $Q$, it follows from equation (\ref{eq:linearYreduced}) and standard properties of the covariance operator that,
\begin{align}
\cov(Q, Y) &= \cov(Q, \mu_Y^\ast) + \beta \, \cov(Q, X) + \psi \, \cov(Q, M) + \\ \nonumber
& \;\;\;\; + \cov(Q, f^\ast(\bfU, \bfmV, \bfmH)) + \cov(Q, \epsilon_Y^\ast) \\ \nonumber
&= \psi \, \cov(Q, M)~,
\end{align}
since $Q \ci \mu_Y^\ast$, $Q \ci X$, $Q \ci f^\ast(\bfU, \bfmV, \bfmH)$, and $Q \ci \epsilon_Y^\ast$, and the respective covariance terms are 0 (here, the symbol $\ci$ stands for statistical independence). Therefore, $\psi$ can be identified as,
\begin{equation}
\psi = \frac{\cov(Q, Y)}{\cov(Q, M)}~,
\label{eq:placeboeffect}
\end{equation}
as long as $\cov(Q, M) \not= 0$ (in practice, this condition is met if the psychological encouragement treatment can effectively manipulate the desire for improved symptoms, which, by its turn influences the emotional state, $M$).

Now, if we let $R = Y - \psi \, M$ represent the residual of the outcome variable, after removal of the placebo effect, then we can rewrite equation (\ref{eq:linearYreduced}) as,
\begin{equation}
R = \mu_Y^\ast + \beta \, X + f^\ast(\bfU, \bfmV, \bfmH) + \epsilon_Y^\ast~.
\label{eq:residualY}
\end{equation}

Because $Z$ is also randomized, it follows from equation (\ref{eq:residualY}) and the properties of the covariance operator that,
\begin{align}
\cov(Z, R) &= \cov(Z, \mu_Y^\ast) + \beta \, \cov(Z, X) + \\ \nonumber
& \;\;\;\; + \cov(Z, f^\ast(\bfU, \bfmV, \bfmH)) + \cov(Z, \epsilon_Y^\ast) \\ \nonumber
&= \beta \, \cov(Z, X)~,
\end{align}
since $Z \ci \mu_Y^\ast$, $Z \ci f^\ast(\bfU, \bfmV, \bfmH)$, and $Z \ci \epsilon_Y^\ast$. Hence, the treatment effect $\beta$ can be identified as,
\begin{equation}
\beta = \frac{\cov(Z, R)}{\cov(Z, X)}~,
\label{eq:treatmenteffect}
\end{equation}
as long as $\cov(Z, X) \not= 0$ (in practice, this condition is met whenever there is some degree of compliance between assigned and received treatments).

Note that in addition to the three core assumptions required by an IV\cite{didelez2010} (namely, that it is statistically independent of any unmeasured confounders; is marginally associated with the treatment variable; and is associated with the outcome variable exclusively through its influence on the treatment variable), the above derivations require that $X$ and $M$ are linearly related to $Y$, but make no assumptions about the relationship between $Y$ and the unmeasured confounders. Estimators for the placebo and treatment effects in equations (\ref{eq:placeboeffect}) and (\ref{eq:treatmenteffect}) are presented in the next subsection.

An alternative statistical framework, based on Rubin's potential outcomes approach to causality\cite{rubin1978,imbensrubin2015}, has been proposed in the literature to address partial compliance in studies involving binary instrumental and treatment variables\cite{angrist1996,imbensrubin2015}. While the method in reference\cite{angrist1996} is not directly applicable to the estimation of the placebo effects, it could be used to estimate treatment effects (after removal of the placebo effect). We point out, however, that the estimator obtained from the potential outcome approach is still identical to the estimator derived from the mechanism based approach, so that statistical inferences based on randomization tests are still the same, independent of the causality framework one is willing to adopt.

\subsection{Two-step estimation procedure}

Adopting a method of moments approach, we have that a consistent estimator for $\psi$ is given by,
\begin{equation}
\hat{\psi} = \frac{\widehat{\cov}(Q, Y)}{\widehat{\cov}(Q, M)} = \frac{\frac{1}{n} \sum_{k=1}^{n} Q_{k} Y_{k} - (\frac{1}{n} \sum_{k=1}^{n} Q_{k}) (\frac{1}{n} \sum_{k=1}^{n} Y_{k})}{\frac{1}{n} \sum_{k=1}^{n} Q_{k} M_{k} - (\frac{1}{n} \sum_{k=1}^{n} Q_{k}) (\frac{1}{n} \sum_{k=1}^{n} M_{k})}~.
\label{eq:placeboIVestimator}
\end{equation}

Note that the above placebo effect estimator requires measurements of $M$, but not of $E$ or $D$. We point out, however, that if expectation and desire measurements are also available, then we can evaluate the validity of the desire-expectation model for the data at hand by checking whether the $E$, $D$, and $I$ variables are able to predict the $M$ measurements. We can also assess the effectiveness of the psychological treatment in influencing desire for better symptoms by estimating $\mbox{Cor}(Q, D)$.

Direct estimation of the treatment effect in equation (\ref{eq:treatmenteffect}) using an IV estimator is unfeasible, as it would involve the unmeasured quantities $R_k = Y - \psi \, M_k$. Therefore, in order to obtain a consistent estimator of the treatment effect, we adopt a two-step approach where we first estimate $R_k$ as $\widehat{R}_{k} = Y_k - \hat{\psi} \, M_k$, for $k = 1, \ldots, n$, and then estimate $\beta$ using,
\begin{equation}
\hat{\beta} = \frac{\widehat{\cov}(Z, \widehat{R})}{\widehat{\cov}(Z, X)} = \frac{\frac{1}{n} \sum_{k=1}^{n} Z_{k} \widehat{R}_{k} - (\frac{1}{n} \sum_{k=1}^{n} Z_{k}) (\frac{1}{n} \sum_{k=1}^{n} \widehat{R}_{k})}{\frac{1}{n} \sum_{k=1}^{n} Z_{k} X_{k} - (\frac{1}{n} \sum_{k=1}^{n} Z_{k}) (\frac{1}{n} \sum_{k=1}^{n} X_{k})}~.
\label{eq:treatmentIVestimator}
\end{equation}

Note that the IV estimators in equations (\ref{eq:placeboIVestimator}) and (\ref{eq:treatmentIVestimator}) can produce highly inflated estimates when $\widehat{\cov}(Q, M) \approx 0$ and $\widehat{\cov}(Z, X) \approx 0$. Hence, in practice, it is important to check the assumptions that psychological encouragement influences the emotion levels, and that the compliance between assigned and received treatments is not negligible.

\subsection{Randomization tests for $H_0: \psi = 0$ and $H_0: \beta = 0$}

We implemented randomization tests\cite{ernst2004} for testing the presence of a placebo effect ($H_0: \psi = 0$ versus $H_1: \psi \not= 0$), and of a treatment effect ($H_0: \beta = 0$ versus $H_1: \beta \not= 0$). The randomization null distribution for the placebo effect is generated by evaluating the statistic $\hat{\psi}$ in equation (\ref{eq:placeboIVestimator}) on a large number of shuffled versions of the data, where the $Y_k$ measurements are shuffled relative to the $(Q_k, M_k)$ measurements (whose connection is kept intact in order to preserve the association between the $Q$ and $M$ variables). The randomization null for treatment effect is generated by first calculating the residuals, $\widehat{R}_{k} = Y_k - \hat{\psi} \, M_k$, where $\hat{\psi}$ is computed in the observed (not permuted) data, and then evaluating the statistic $\hat{\beta}$ in equation (\ref{eq:treatmentIVestimator}) in shuffled data sets, where $R_k$ is shuffled relative to $(Z_k, X_k)$ data (whose connection is kept intact to preserve the association between $Z$ and $X$). These randomization tests are non-parametric procedures and don't make any distributional assumptions about the data. However, because the identification of the causal effects assumes a linear relation between $Y$ and $X$ and $M$, the validity of the tests is still contingent on this assumption.

\subsection{Randomization confidence intervals}

Here we describe how to build confidence intervals for placebo and treatment effects using the p-values from randomization tests\cite{garthwaite1996,ernst2004}. Throughout this section we use $\theta$ to represent either the placebo effect, $\psi$, or the treatment effect, $\beta$. The procedure is strait-forward but requires a considerable amount of computation (which, nonetheless, can be easily parallelized). Assume for a moment that randomization tests for testing $H_0: \theta = \theta_j$ against one-sided alternative hypotheses $H_1: \theta < \theta_j$ and $H_1: \theta > \theta_j$ are available. Exploring the correspondence between confidence intervals and hypothesis tests, we obtain a $100 (1 - 2 \alpha)$\% confidence interval (CI) for $\theta$ by searching for a lower bound value, $\theta_L$, such that $H_0: \theta = \theta_L$ is rejected in favor of $H_1: \theta > \theta_L$ at a significance $\alpha$, and by searching for an upper bound value, $\theta_U$, such that $H_0: \theta = \theta_U$ is rejected in favor of $H_1: \theta < \theta_U$ at the same significance level\cite{garthwaite1996}.

While an efficient algorithm for finding CI bounds has been proposed\cite{garthwaite1996}, the approach requires the specification of the significant level before hand. In order to avoid this constraint, we generate a one-sided randomization p-value profile which can be used to determine the $100 (1 - 2 \alpha)$\% CI for any desired $\alpha$ level. This p-value profile is generated as follows: ($i$) compute the observed placebo or treatment effect estimate, $\hat{\theta}$; ($ii$) for each $\theta_j < \hat{\theta}$, in a grid of decreasing $\theta_j$ values, compute the randomization p-value from the one-sided test $H_0: \theta = \theta_j$ vs $H_1: \theta > \theta_j$; ($iii$) repeat step $ii$ until a p-value equal to zero is reached; ($iv$) for each $\theta_j > \hat{\theta}$, in a grid of increasing $\theta_j$ values, compute the p-value from the one-sided test $H_0: \theta = \theta_j$ vs $H_1: \theta < \theta_j$; ($v$) repeat step $iv$ until a randomization p-value equal to zero is found.

Before we explain how to generate null distributions for placebo effects different from zero, consider first the intention-to-treat (ITT) estimator,
\begin{equation}
\widehat{ITT}_\psi = \frac{\sum_{k=1}^{n} Y_k \, \ind\{ Q_k = 1 \}}{\sum_{k=1}^{n} \ind\{ Q_k = 1 \}} - \frac{\sum_{k=1}^{n} Y_k \, \ind\{ Q_k = 0 \}}{\sum_{k=1}^{n} \ind\{ Q_k = 0\}} = \frac{\widehat{\cov}(Q, Y)}{\widehat{\mbox{Var}}(Q)}~.
\end{equation}
Instead of directly generating a randomization distribution under the null $H_0: \psi = \psi_j$, we generate a randomization distribution under the equivalent null hypothesis that the intention-to-treat effect is equal to $\psi_j \, K_1$, where $K_1 = \widehat{\cov}(Q, M)/\widehat{\mbox{Var}}(Q)$ is constant across all permutations of the response data used in the construction of the randomization null. (Note that, because $\widehat{ITT}_\psi = K_1 \, \hat{\psi}$ the randomization tests based on $\hat{\psi}$ and $\widehat{ITT}_\psi$ estimators produce exactly the same p-value if we use the same permutations of the response data in the construction of their null distributions.)

The practical advantage of the test based on $\mbox{ITT}_\psi$ effects is that it amounts to a simple two sample location problem for testing whether the difference in average response between the assigned treatment (psychological encouragement) and assigned control (no encouragement) groups is equal to $\psi_j \, K_1$. The implementation of randomization tests for this two sample location problem is strait-forward\cite{garthwaite1996}: we only need to add $\psi_j \, K_1$ for each $Y_k$ data point in the assigned control group (i.e., $k$ for which $Q_k = 0$), while leaving the response data from the assigned treatment group, $Q_k = 1$, unchanged, and then run a randomization test for testing the null hypothesis that the $\mbox{ITT}_\psi$ effect is equal to zero, against the alternative one-sided hypothesis that it is positive, and against the alternative that it is negative.

Similarly, for the treatment effects we consider the two-step ITT estimator,
\begin{equation}
\widehat{ITT}_\beta = \frac{\sum_{k=1}^{n} \hat{R}_k \, \ind\{ Z_k = 1 \}}{\sum_{k=1}^{n} \ind\{ Z_k = 1 \}} - \frac{\sum_{k=1}^{n} \hat{R}_k \, \ind\{ Z_k = 0 \}}{\sum_{k=1}^{n} \ind\{ Z_k = 0\}} = \frac{\widehat{\cov}(Z, \hat{R})}{\widehat{\mbox{Var}}(Z)}~,
\end{equation}
and generate randomization distributions under the equivalent null hypotheses $H_0: \mbox{ITT}_\beta = \beta_j \, K_2$, where $K_2 = \widehat{\cov}(Z, X)/\widehat{\mbox{Var}}(Z)$, by simply adding $\beta_j \, K_2$ for each $\hat{R}_k$ data point in the assigned control group, $Z_k = 0$, while leaving the residual data from the assigned treatment group, $Z_k = 1$, unchanged (and then testing for the null that the $\mbox{ITT}_\beta$ is equal to zero, against the alternative one-sided hypotheses that it is positive and the alternative that it is negative0.

\subsection{Adjustment for observed confounders}

If measured confounders influencing both $X$ and $Y$ are available, it is possible to adjust for them by simply working with the residuals of $X$ and $Y$ (computed by separately regressing $X$ and $Y$ on the confounders). Similarly, if measured confounders influencing both $M$ and $Y$ are available, it is possible to adjust for them by working with the respective residuals.

\subsection{Regression based estimators and tests}

We compare the proposed IV estimators, and their respective randomization tests, to standard estimators and analytical hypothesis tests based on the linear regression of the outcome variable, $Y$, on both the received treatment, $X$, and emotion level, $M$, according to the model, $Y = \mu_Y + \beta \, X + \psi \, M + \epsilon_Y$. Under this regression based approach, we estimate $\beta$ and $\psi$ using ordinary least squares, and test the null hypotheses $H_0: \psi = 0$ and $H_0: \beta = 0$ using standard t-tests. In our simulations (described next), we generate data using gaussian errors, so that the distributional assumptions underlying the analytical t-tests are met.

\subsection{Simulation experiments details}

We simulated data from blinded and unblinded settings, in the presence or absence of confounding, according to the models presented in Figure \ref{fig:simdags}. For each of these settings, we run 4 separate simulation studies generating data: (i) under the null hypothesis that both treatment and placebo effect are zero, $H_0: \beta = 0$ and $H_0: \psi = 0$; (ii) under the alternative for treatment effects, $H_1: \beta \not= 0$, but null for placebo effects, $H_0: \psi = 0$; (iii) the other way around, $H_0: \beta = 0$ and $H_1: \psi \not= 0$; and (iv) under the alternative for both treatment and placebo effects, $H_1: \beta \not= 0$ and $H_1: \psi \not= 0$.

Each simulated data set was generated as follows. The IVs $Z$ and $Q$ were sampled from $\mbox{Bernoulli}(1/2)$ distributions. All confounding variables were sampled from $\mbox{Normal}(0, 1)$ distributions. The binary variables $X$, $E$, and $D$ were generated by the threshold models,
\begin{align}
X &= \ind\{ \theta_{XZ} \, Z + \theta_{XU} \, U + \theta_{XC_1} \, C_1 + \theta_{XC_2} \, C_2 + \theta_{XC_3} \, C_3 + \epsilon_{X} > 0 \}~, \\
E &= \ind\{ \theta_{EX} \, X + \theta_{EC_1} \, C_1 + \theta_{EL_1} \, L_1 + \theta_{EV_2} \, V_2 + \theta_{EL_3} \, L_3 + \epsilon_{E} > 0 \}~, \\
D &= \ind\{ \theta_{DQ} \, Q + \theta_{DV_1} \, V_1 + \theta_{DC_2} \, C_2 + \theta_{DL_2} \, L_2 + \theta_{DL_3} \, L_3 + \epsilon_{D} > 0 \}~,
\end{align}
where $\epsilon_{X}$, $\epsilon_{E}$, and $\epsilon_{D}$ were sampled from $\mbox{Normal}(0, 1)$ distributions. The interaction $I$ was generated as the product of $E$ and $D$. Finally, the emotion and outcome data were generated from the linear models,
\begin{equation}
M = \theta_{ME} \, E + \theta_{MD} \, D +  \theta_{MI} \, I +  \theta_{ML_1} \, L_1 +  \theta_{ML_2} \, L_2 +  \theta_{MC_3} \, C_3 +  \theta_{MV_3} \, V_3 + \epsilon_{M}~,
\vspace{-0.3cm}
\end{equation}
\begin{equation}
Y = \theta_{YX} \, X + \theta_{YM} \, M + \theta_{YU} \, U + \theta_{YV_1} \, V_1 + \theta_{YV_2} \, V_2 + \theta_{YV_3} \, V_3 + \epsilon_{Y}~,
\end{equation}
where $\epsilon_{M}$ and $\epsilon_{Y}$ were sampled from $\mbox{Normal}(0, 1)$ distributions. (Note that the explicit form of the desire-expectation model of emotions is unimportant, as the estimator for $\psi$ depends on the observed values of $M$, but not of $E$, $D$, and $I$, and does not require an explicit description of the functional relationships between $M$ and $E$, $D$, and $I$. Hence, for simplicity, we adopt a simple linear relation, even though more sophisticated relations could have been used.)

Each simulation experiment comprised 10,000 distinct synthetic data sets. Each simulated data set was generated using a unique combination of simulation parameter values. In order to select parameter values spread as uniformly as possible over the entire parameter range we employed a Latin hypercube design\cite{santner2003}, optimized according to the maximin distance criterium \cite{johnson1990}, in the determination of the parameter values used on each of the 10,000 simulated data sets for each simulation experiment.

We selected wide ranges for all model parameters. Explicitly, the parameters representing the effect of confounders on the observed variables (namely, $\theta_{XU}$, $\theta_{XC_1}$, $\theta_{XC_2}$, $\theta_{XC_3}$, $\theta_{EC_1}$, $\theta_{EL_1}$, $\theta_{EV_2}$, $\theta_{EL_3}$, $\theta_{DV_1}$, $\theta_{DC_2}$, $\theta_{DL_2}$, $\theta_{DL_3}$, $\theta_{ML_1}$, $\theta_{ML_2}$, $\theta_{MC_3}$, $\theta_{MV_3}$, $\theta_{YU}$, $\theta_{YV_1}$, $\theta_{YV_2}$, and $\theta_{YV_3}$) were selected in the range $[-2, 2]$ for the simulations under the influence of confounders, but were set to 0 in the simulations under unconfounded conditions. The effect of $Z$ on $X$ ($\theta_{XZ}$), and of $Q$ on $D$ ($\theta_{DQ}$), as well as, the effects of $E$, $D$, and $I$ on $M$ ($\theta_{ME}$, $\theta_{MD}$, and $\theta_{MI}$) were selected in the range $[1, 2]$. The effect of $X$ on $E$ ($\theta_{EX}$) was set to 0 in the blinded setting simulations, and selected in the range $[1, 2]$ in the unblinded simulations. The treatment effect ($\beta$) and the placebo effect ($\psi$) parameters were set to 0 in the simulations under the null hypothesis, and were selected in the range $[-2, 2]$ for the simulations under the alternative hypothesis. The range of sample size parameter, $n$, was set to realistic values we expect to see in practice, $\{100, 101, \ldots, 1000\}$.

For any fixed significance level $\alpha$, the empirical type I error rate was computed as the fraction of the p-values smaller than $\alpha$ over the data sets simulated under the null hypothesis, whereas the empirical power was calculated as the fraction of p-value smaller than $\alpha$ over data sets generated under the alternative hypothesis.



\section{Acknowledgements}

This work was funded by the Robert Wood Johnson Foundation.

\end{document}